\documentclass[twocolumn,           
               showpacs,            
               preprintnumbers,     
               aps,                 
               prd,          	    
               a4paper,             
               groupedaddress,      
               nofootinbib,         
               tightenlines,        
               floats,floatfix      
               ]{revtex4}

\usepackage{graphicx}
\usepackage{latexsym}
\usepackage{amsmath,amssymb}        
\usepackage[draft=false]{hyperref}

\newcommand{\de}{\mathrm{d}}
\renewcommand{\(}{\left(}
\renewcommand{\)}{\right)}
\renewcommand{\[}{\left[}
\renewcommand{\]}{\right]}
\newcommand{\period}{\,\mathrm{.}}
\newcommand{\comma}{\,\mathrm{,}}
\newcommand{\reffig}[1]{Fig.~\ref{#1}}
\newcommand{\refeq}[1]{Eq.~(\ref{#1})}
\newcommand{\mpl}{m_\mathrm{Pl}}

\newcommand{\abs}[1]{\left\vert#1\right\vert}
\newcommand{\pd}{\partial}

\begin{document}



\title{Initial conditions for quintessence after inflation}
\author{Micha\"{e}l Malquarti}
\author{Andrew R.~Liddle}
\affiliation{Astronomy Centre, University of Sussex, Brighton BN1 9QJ, United 
Kingdom}
\date{\today}
\pacs{98.80.Cq \hfill astro-ph/0203232}
\preprint{astro-ph/0203232}


\begin{abstract}
We consider the behaviour of a quintessence field during an inflationary epoch, 
in order to learn how inflation influences the likely initial conditions for 
quintessence. We use the stochastic inflation formalism to study quantum 
fluctuations induced in the quintessence field during the early stages of 
inflation, and conclude that these drive its mean to large values ($\gtrsim 0.1 
\mpl$). Consequently we find that tracker behaviour typically starts at low 
redshift, long after nucleosynthesis and most likely also after decoupling.
\end{abstract}

\maketitle


\section{Introduction}

In order to fit the impressive array of observational data now available, it is 
widely believed that a model of the Universe should feature a period of 
inflation in its early history, leading to the generation of density 
perturbations via quantum fluctuations, and that the present expansion rate 
should be accelerating. While the latter phenomenon is often considered due to
a cosmological constant, it is at least as attractive to presume that the 
acceleration is driven by the same mechanism usually exploited to give early 
Universe inflation, namely domination by the potential energy of a scalar
field. Such models are known as quintessence models. 

An important class of quintessence models are known as tracking models
\cite{track,RP,nucl2,scal,equip}, where the late-time evolution of the field
has an attractor behaviour rendering it fairly independent of initial
conditions.
However, despite the existence of tracking behaviour the details of the initial
conditions may yet be important.  For example, in several models of
quintessence the tracking solution cannot be achieved too early; big bang
nucleosynthesis is spoiled if the quintessence field has too large a density at
that time \cite{nucl2,nucleo}.  More obviously, a quintessence scenario will
not work unless the initial energy density is at least as large as is required 
by the present.  It is therefore useful to have further guidance as to the 
likely initial conditions for quintessence.

The simplest assumption concerning quintessence is that it is a fundamental 
scalar field (rather than a low-energy composite field), and as such was already 
present during the inflationary epoch in the early Universe. If the quintessence 
field is sufficiently weakly coupled that it is not affected by the inflaton 
decays ending inflation, its possible initial conditions are restricted by its 
dynamics during the inflationary period. Note that we are not considering the 
situation where the quintessence field and the inflaton are the same field 
\cite{quintinf}.

In this paper we carry out a comprehensive study of the influence of 
inflationary dynamics on the quintessence field. We choose parameters so that 
the quintessence sector matches the observed acceleration of the Universe, and 
the inflaton sector generates suitable perturbations to initiate structure 
formation. We investigate both classical and quantum dynamics of the 
quintessence. We assume a flat Universe throughout.

\section{Models and observational constraints}

\subsection{Quintessence}

The quintessence model is defined by a scalar field $Q$ evolving in a potential 
$V(Q)$. Constraints on such models from observation have recently been 
considered by various authors \cite{obscon,BHM02}; the precise details are not 
important for our considerations but for definiteness we use the results of
Bean et al.~\cite{BHM02}, who recently combined constraints on such models 
from type Ia supernovae, CMB peak positions and large-scale structure surveys. 
They found one-sigma constraints on the quintessence density parameter and its 
pressure--density ratio of 
\begin{equation}
0.57\leq\Omega_Q\leq0.72\quad\mathrm{and}\quad
-1\leq w_Q\leq-0.85\period
\end{equation}
With this, and assuming the Hubble constant to be given by $h=0.72\pm0.08$ 
\cite{HKP}, it is possible to find constraints on the parameters of the 
quintessence models. To locate viable regions of parameter space, we approximate 
these three constraints as Gaussian distributed (in the case of $w_Q$ a 
half-Gaussian centred at $w_Q = -1$) and independent. This defines a confidence 
region in the space of parameters, and using numerical simulations we can 
translate the confidence region into other variables such as the density of 
non-relativistic matter today and the parameters of the quintessence potential.

\begin{figure}[t]
\includegraphics[scale=0.37,angle=-90]{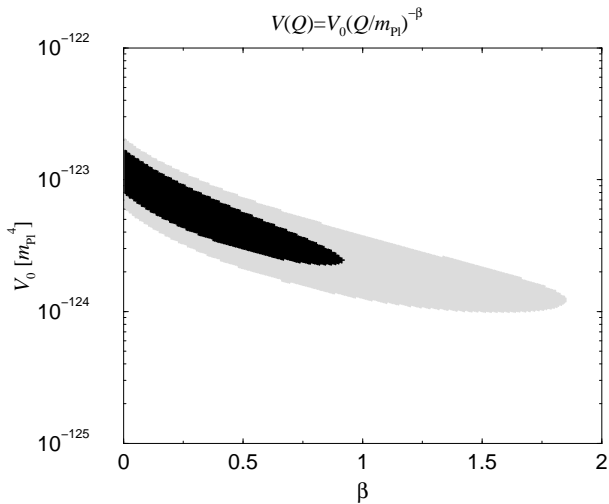}
\caption{The 68\% and 95\% confidence regions for an inverse power-law model of
quintessence.}\label{conf}
\end{figure}

Our main discussion will focus on quintessence models with inverse power-law 
potentials, as originally introduced by Ratra and Peebles \cite{RP}
\begin{equation}
V(Q)=V_0\(\frac{Q}{\mpl}\)^{-\beta}\period
\end{equation}
In order to simplify some later expressions we define the notation 
$\tilde{X}=X/\mpl$, where $X$ is any field or variable. Using the procedure 
described above and projecting the result onto the $\beta$--$V_0$ plane gives 
the allowed region shown in \reffig{conf}. We see that $\beta\lesssim1$ and 
$V_0\simeq5\times10^{-124} \mpl^4$ at 68\% confidence, with a broader region at  
95\% confidence. In defining our models, we allow $\beta$ to vary and for each 
choice fix $V_0$ at its best-fit value. 

\subsection{Inflation}

We take inflation to be driven by a scalar field $\phi$ (the inflaton) in a
potential $W(\phi)$. Because of the small value of $V_0$, the inflaton dominates 
the evolution and therefore we can use the standard constraints on its 
parameters. The inflation model is to be chosen so as to reproduce the amplitude 
of perturbations needed to generate the observed structures, which requires 
\cite{LL}
\begin{equation}
\frac{512\pi}{75}\,\frac{W^3}{\mpl^6\left|\de W/\de\phi\right|^2}\simeq 
4\times 10^{-10}\period
\end{equation}
This is to be evaluated at $\phi_\mathrm{obs}$, the value of the field
corresponding to the epoch when the scale of the observable Universe crossed
the horizon. 

In this paper we will consider potentials where inflation ends by violation of 
the slow-roll conditions, so the slow-roll approximation
\begin{align}
\varepsilon(\phi) &= \frac{\mpl^2}{16\pi}\(\frac{\de W/\de\phi}{W}\)^2\ll1\\
\eta(\phi) &= \frac{\mpl^2}{8\pi}\frac{\de^2W/\de\phi^2}{W}\ll1
\end{align}
gives the end of inflation and the number of $e$-foldings
\begin{equation}
N\simeq\frac{8\pi}{\mpl^2}
\int^{\phi_\mathrm{obs}}_{\phi_\mathrm{end}}\frac{W(\phi)}{\de W/\de\phi}
\,\de\phi\simeq50
\end{equation}
gives us $\phi_\mathrm{obs}$. 

Fifty or so $e$-foldings is the minimum amount of inflation capable of giving 
our observed Universe, but crucial for the considerations of this paper is that 
typically one expects enormously more inflation. If one only imposes the 
constraint that the initial energy density be below the Planck energy, most 
models allow very large amounts of inflation. However the early stages of such 
inflation may be dominated by quantum dynamics (the so-called stochastic 
inflation regime \cite{SI}). The maximum value of the inflaton field we 
consider, $\phi_\mathrm{max}$, is given by the quantum-to-classical evolution 
transition
\begin{equation}
\frac{\Delta_\mathrm{qu}}{\Delta_\mathrm{cl}}=
\frac{H}{2\pi}\abs{\frac{H}{\dot{\phi}}}\lesssim1\,,
\end{equation}
where the $\Delta$ represent the typical quantum and classical evolution per 
Hubble time.

For simplicity we focus on a inflaton model with a power-law potential
\begin{equation}
W(\phi)=W_0\(\frac{\phi}{\mpl}\)^{\alpha}\period
\end{equation}
The former constraints lead to
\begin{eqnarray}
\tilde{\phi}_\mathrm{end}&\simeq& \mathrm{Max}\[\frac{\alpha}{4\sqrt{\pi}}
\,\mathrm{,}\,
\sqrt{\frac{\alpha(\alpha-1)}{8\pi}}\;\]\comma\\
\tilde{\phi}_\mathrm{obs}&\simeq& 
\sqrt{\frac{50\alpha}{4\pi}+\tilde{\phi}_\mathrm{end}^2}\comma\\
\tilde{\phi}_\mathrm{max}&\simeq&
\(\frac{3\alpha^2\mpl^4}{128\pi W_0}\)^{1/(\alpha+2)} \comma \\
W_0&\simeq&\frac{3\times10^{-8}\alpha^2\mpl^4}{512\pi}\,
         \tilde{\phi}_\mathrm{obs}^{-(\alpha+2)}\period
\end{eqnarray}

\section{Quintessence field evolution during inflation}

The models are now fully defined and we can study the dynamics of each field, 
beginning with the classical evolution.

\subsection{Classical evolution}

We assume that the quintessence field is completely uncoupled from everything 
else, including the inflaton. The equations of motion are the
Euler--Lagrange equations for the inflaton and quintessence fields
\begin{equation}
\ddot{\phi}=-3H\dot{\phi}-\frac{\de W}{\de\phi}
\quad\mathrm{;}\quad
\ddot{Q}=-3H\dot{Q}-\frac{\de V}{\de Q}
\end{equation}
and the Einstein equation
\begin{equation}
H^2=\frac{8\pi}{3 \mpl^2} \left[ \frac{1}{2} \dot{\phi}^2+W(\phi)+
	\frac{1}{2} \dot{Q}^2+V(Q) \right] \period
\end{equation}
Because of the very small value of $V_0$, the evolution of the Universe is
determined by the inflaton. The quintessence field may dominate if its value is 
very small, but, as we will see this situation would end very quickly. 
Therefore, using the slow-roll approximation we have the solution
\begin{equation}
\tilde{\phi}(N)\simeq\sqrt{\tilde{\phi}^2_\mathrm{ini}
-\frac{\alpha N}{4\pi}}\comma
\end{equation}
where $N=\log(a/a_\mathrm{ini})$. The subscript ``ini'' always means the value 
at some time $t_\mathrm{ini}$. Knowing the dynamics of the Universe we can now 
study the quintessence field.

The classical behaviour of the quintessence field is rather straightforward. If 
the field is in a region where $\abs{\de V/\de Q}$ is small, its classical 
evolution will be highly suppressed by the friction arising from the 
inflationary expansion and to a good approximation the field will retain its 
initial value. Conversely, if $\abs{\de V/\de Q}$ is large, the field will 
quickly roll down until the potential becomes too flat. More precisely, using 
the slow-roll approximation for the quintessence field we have
\begin{equation}
\dot{\tilde{Q}}\simeq\frac{\beta\,V_0}{3H(t)\mpl^2}\,\tilde{Q}^{-\beta-1}
\period
\end{equation}
Assuming $H(t)$ constant (roughly correct for a few Hubble times) the solution 
is
\begin{equation}
\tilde{Q}(t)\simeq\left[ \frac{\beta(\beta+2)V_0}{3H\mpl^2}(t-t_\mathrm{ini})
+\tilde{Q}_\mathrm{ini}^{\beta+2}\right]^{1/(\beta+2)}\period
\end{equation}
We define
\begin{equation}
\tilde{Q}_\mathrm{min}=\left[\frac{\beta(\beta+2)V_0}{3H^2\mpl^2}
	\right]^{1/(\beta+2)}
\end{equation}
and see that the evolution during a Hubble time has the following
approximate description:
\begin{equation}
\begin{array}{rll}
\tilde{Q}(t_\mathrm{ini}+H^{-1})&\simeq\tilde{Q}_\mathrm{ini}&
\quad\mathrm{if}\quad \tilde{Q}_\mathrm{ini}\gg \tilde{Q}_\mathrm{min}\\
\tilde{Q}(t_\mathrm{ini}+H^{-1})&\simeq\tilde{Q}_\mathrm{min}&
\quad\mathrm{if}\quad \tilde{Q}_\mathrm{ini}\ll \tilde{Q}_\mathrm{min}
\end{array}
\period
\end{equation}
Therefore, after a few Hubble times the classical evolution of the quintessence
field is to remain constant.

\subsection{Quantum fluctuations}

Although the large friction from the Hubble expansion renders the classical 
evolution of the quintessence field negligible, the same is not necessarily true 
of the quantum fluctuations. Indeed, as we shall see, the effect of these in the 
quintessence potential is to drive typical regions of the Universe to quite 
large values of the quintessence field, corresponding to low energy densities. 
This possibility was first mentioned in Ref.~\cite{SS}.

Let us now recall briefly the way we treat these fluctuations. Following 
Ref.~\cite{LL}, we choose the spatially-flat slicing and we split the 
quintessence 
field into an unperturbed part and a perturbation
\begin{equation}
Q(\mathbf{x},t)=Q(t)+\delta Q(\mathbf{x},t)\period
\end{equation}
We quantize the perturbation and expand its Fourier components
\begin{equation}
\delta \hat{Q}_\mathbf{k}(t)=w_k(t)\hat{a}_\mathbf{k}
                      +w^*_k(t)\hat{a}^\dagger_{-\mathbf{k}} \comma
\end{equation}
where $\hat{a}_\mathbf{k}$ is the annihilation operator. In the linear 
approximation we have to solve
\begin{equation}\label{fieldeq}
\ddot{w}_k+3H\dot{w}_k+\[(k/a)^2+m_\mathrm{eff}^2\]w_k=0\comma
\end{equation}
where $m_\mathrm{eff}^2=\de^2V/\de Q^2$. If the 
field is effectively massless and $H$ is approximately constant, the power 
spectrum of the fluctuations is given by
\begin{equation}
\mathcal{P}_k(t)=\frac{L^3k^3}{2\pi^2}\abs{w_k(t)}^2
                \simeq\[1+\(\frac{k}{aH}\)^2\]\(\frac{H}{2\pi}\)^2\period
\end{equation}
This means that before horizon-crossing $\mathcal{P}_k(t)\propto 
1/a^2(t)$ and afterwards it becomes constant. Using
\begin{equation}
\langle\delta\hat{Q}^2(\mathbf{x},t)\rangle
=\int^\infty_0\mathcal{P}_k(t)\frac{\de k}{k}\comma
\end{equation}
and the fact that the fluctuations become classical after they cross the 
horizon, we find that the quintessence field receives quantum kicks whose size 
is estimated as $H/2\pi$ per Hubble time.

The effect of these quantum fluctuations can be studied using the Fokker--Planck 
formalism \cite{SI2}, which allows one to follow the probability distribution 
$f(Q,t)$ of the quintessence field during inflation. A simple derivation is as 
follows. We take discrete steps $\Delta t = H^{-1}\varepsilon$
during which there are a random jump of magnitude
$\Delta Q_\mathrm{qu}=\sqrt{\Delta t/H^{-1}}H/2\pi
                     =\sqrt{\Delta \varepsilon}H/2\pi$
due to quantum fluctuations and a classical step
$\Delta Q _\mathrm{cl}=\dot{Q}\Delta t
                      =-\varepsilon V^{\prime}/3H^2$,
where the prime denotes a derivative with respect to $Q$. During this time an 
interval $\de Q$ shrinks by a factor $1+\varepsilon V^{\prime\prime}/3H^2$. 
We find the following equation:
\begin{equation}
\begin{split}
f(Q,t+\Delta t)\de Q =&
\,\frac{1}{2}\Big[f(Q-\Delta Q_\mathrm{cl}+\Delta Q_\mathrm{qu},t)\\
&+f(Q-\Delta Q_\mathrm{cl}-\Delta Q_\mathrm{qu},t)\Big]\\
&\times(1+\varepsilon V^{\prime\prime}/3H^2)\, \de Q \period
\end{split}
\end{equation}
Subtracting $f(Q,t)\de Q$, dividing by $\varepsilon$ and taking the limit
$\varepsilon\to0$ leads to
\begin{equation}
\begin{split}
\frac{\pd f(Q,t)}{\pd t}=&
\,\frac{H^{3}(t)}{8\pi^2}\,\frac{\pd^2 f(Q,t)}{\pd Q^2}\\
&+\frac{1}{3H(t)}\frac{\pd}{\pd Q}\(\frac{\pd V(Q)}{\pd Q}\,f(Q,t)\)\period
\end{split}
\end{equation}
On the right-hand side, the first term produces diffusion and the second one 
produces a drift. Note that unlike the usual stochastic inflation situation, 
here $H(t)$ is determined externally by the inflaton field evolution, and the 
fluctuations in $Q$ do not back-react on the expansion rate.

To be complete we need the boundary condition
\begin{equation}
\lim_{Q\to0}\left[\frac{H^3(t)}{8\pi^2}\frac{\pd f(Q,t)}{\pd Q}
+\frac{\pd V(Q)/\pd Q}{3H(t)}f(Q,t)\right]=0\comma
\end{equation}
the requirement that $f(Q,t)\geq0$, and normalization $\int_0^\infty f(Q,t)\de 
Q=1$ which implies $\lim_{Q\to\infty}f(Q,t)=0$.

{}From now on we focus on the inverse power-law potential for the quintessence. 
First of all, one may wonder whether the massless condition we used for the 
typical quantum jump is justified. Indeed, for values smaller than 
$Q_\mathrm{lim}$, where
\begin{equation}
\tilde{Q}_\mathrm{lim}=
\left[\frac{\beta(\beta+1)V_0}{H^2\mpl^2}\right]^{1/(\beta+2)}
\comma
\end{equation}
the quintessence field has an effective mass $m_\mathrm{eff}$ which is no longer 
negligible compared to the Hubble rate.\footnote{For power-law potentials, this 
field value is almost the same as $Q_{{\rm min}}$, the value below 
which the classical evolution is important.}
For all wavelengths $a/k$ bigger than $1/m_\mathrm{eff}$, 
\refeq{fieldeq} implies
\begin{equation}
w_k(t)\propto a^{-3/2}(t)
\end{equation}
and therefore the power spectrum of $P_k(t)$ decreases as $1/a^3(t)$ leading to 
a smaller value of the quantum jump after horizon-crossing. Nevertheless, 
because the typical quantum jump $H/2\pi$ (typically $10^{-5}\mpl$) is 
dramatically bigger than $Q_\mathrm{lim}$ (with $\beta=1$, typically 
$10^{-39}\mpl$), the massless condition is broken only during very short periods 
of time (much less than a Hubble time) and so the massless approximation is an 
excellent one.

The Fokker--Planck equation becomes
\begin{equation}\label{eqfp}
\begin{split}
\frac{\pd f(Q,t)}{\pd t}=&
\,\frac{H^{3}(t)}{8\pi^2}\,\frac{\pd^2 f(Q,t)}{\pd Q^2}\\
&-\frac{\pd}{\pd Q}\(\frac{\beta 
V_0\mpl^\beta}{3H(t)}Q^{-\beta-1}\,f(Q,t)\)\period
\end{split}
\end{equation}
For $Q>Q_\mathrm{min}$ we can drop the terms coming from the classical evolution 
since we have seen it is negligible. For $Q<Q_\mathrm{min}$ the effect of the 
potential becomes important and it acts as a wall preventing the quantum 
fluctuations driving the quintessence field to the origin. Therefore, to a 
good approximation we have to solve
\begin{equation}
\frac{\pd f(Q,t)}{\pd t}=
\frac{H^{3}(t)}{8\pi^2} \, \frac{\pd^2 f(Q,t)}{\pd Q^2}\comma
\end{equation}
with the boundary condition $\de f(Q\approx0,t)/\de Q=0$ (no flux 
through the origin) and maintaining $f(Q,t)\geq0$ and $\int_0^\infty f(Q,t)\de 
Q=1$. This is equivalent to a random walk with a wall. Starting with a (half)
Gaussian distribution centred on $Q=0$ and with variance $\sigma^2(t)$, and 
assuming $H(t)$ is constant during $\Delta t$, we have
\begin{equation}
\sigma^2(t_\mathrm{ini}+\Delta t)=
\sigma_\mathrm{ini}^2+\frac{H^3}{4\pi^2}\,\Delta t\period
\end{equation} 
More precisely, the equation to solve is
\begin{equation}
\frac{\de\sigma^2}{\de t}=\frac{H^3(t)}{4\pi^2}\comma
\end{equation}
which has solution 
\begin{equation}
\sigma^2(\tilde{\phi})\simeq\sigma^2_\mathrm{ini}
+\frac{16W_0}{3\alpha(\alpha+2)\mpl^2}
\(\tilde{\phi}_\mathrm{ini}^{\alpha+2}-\tilde{\phi}^{\alpha+2}\)\period
\end{equation}
After enough time the evolution is roughly 
independent of the initial condition $\sigma_\mathrm{ini}$ and we can set it to 
zero.

\begin{figure}[t]
\includegraphics[scale=0.38,angle=-90]{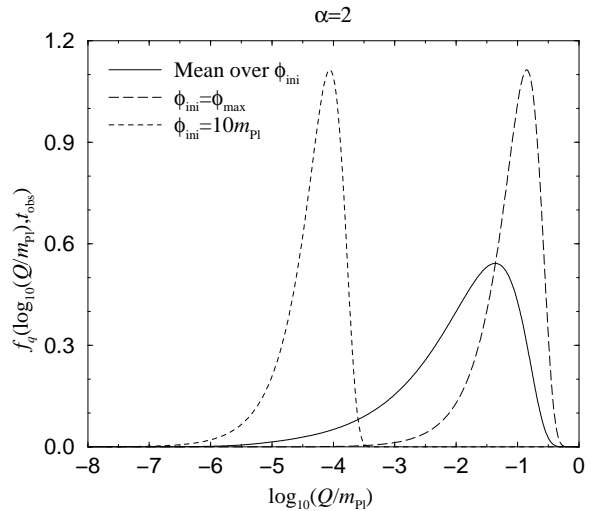}
\caption{Probability distributions for the mean quintessence
value at the end of inflation for different initial values $\phi_\mathrm{ini}$, 
in the particular case of $\alpha=2$. We have also plotted the
probability distribution obtained by averaging over 
$\phi_\mathrm{ini}$.}\label{q_obs}
\end{figure}

We can now ask what one expects the mean value of the quintessence field to be 
in our region of the Universe. Because we are presently interested only in the 
mean value, we should only consider perturbations on scales larger than our 
present horizon, which are generated between the initial inflaton value 
$\phi_\mathrm{ini}$ and $\phi_\mathrm{obs}$. Because of the presence 
of the wall, the net effect of the quantum fluctuations is to diffuse the 
quintessence field to larger values. The extent of this diffusion depends on how 
much inflation occurs before the last fifty $e$-foldings; if more inflation 
occurs then there are more `steps' in the diffusion and also the early steps are 
larger. We have also computed the distribution obtained by averaging over all 
possible initial inflaton values between $\phi_\mathrm{obs}$ and 
$\phi_\mathrm{max}$ assuming a flat probability.

\begin{figure}[t]
\includegraphics[scale=0.38,angle=-90]{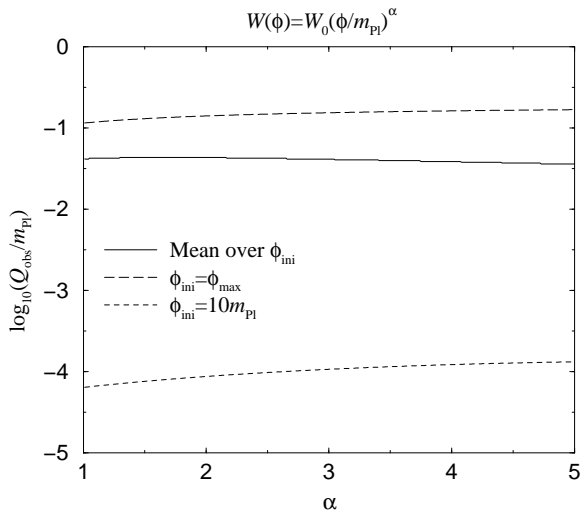}
\caption{The position of the peak of the probability distributions of $Q$, as in 
\reffig{q_obs}, plotted as a function of $\alpha$.}\label{peak}
\end{figure}

Actually we are more interested in the probability distribution $f_q$ of 
$q=\log_{10}(Q/\mpl)$. We have $f_q(q,t)=f(Q(q),t)\de Q/\de q$ which has a 
peak at $q_\mathrm{peak}=\log_{10}(\sigma(t)/\mpl)$. In \reffig{q_obs} we show 
some distributions $f_q(q,t_\mathrm{obs})$ in the particular case of $\alpha=2$; 
we show the probability distributions for two possible initial conditions for 
the inflaton, and then the probability distribution averaged over a uniform 
initial distribution for $\phi_\mathrm{ini}$ between $\phi_\mathrm{obs}$ and 
$\phi_\mathrm{max}$. \reffig{peak} shows the position of the peak in the 
distributions as a function of $\alpha$; as we can see, the result is roughly 
independent of $\alpha$, and as expected the more inflation there is the further 
the distribution diffuses to large $Q$.

\begin{figure}[t]
\includegraphics[scale=0.38,angle=-90]{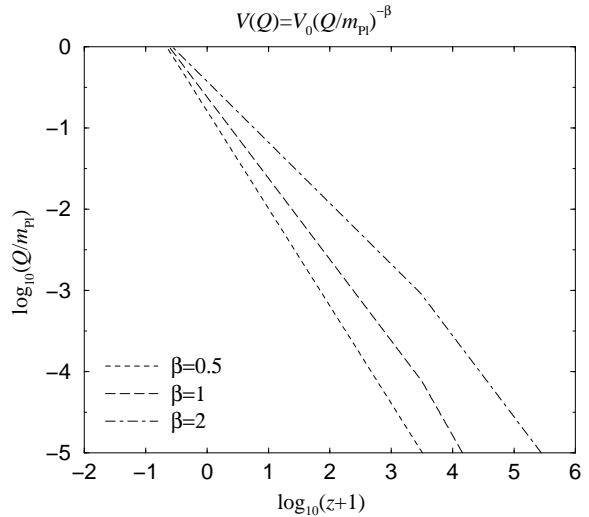}
\caption{Some examples of trackers for different values of 
$\beta$.}\label{tracker}
\end{figure}

\begin{figure}[t!]
\includegraphics[scale=0.38,angle=-90]{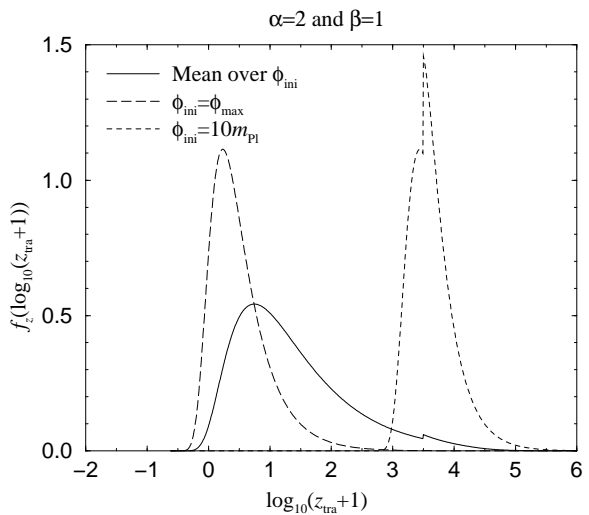}
\caption{Probability distributions for the redshift at which the quintessence
field reaches the tracker, for $\alpha=2$ and $\beta=1$. The distributions are 
shown for two different initial values $\phi_\mathrm{ini}$, and averaged over
$\phi_\mathrm{ini}$. The discontinuities arise across matter--radiation 
equality.}\label{redshift}
\end{figure}

We wish to know the quintessence value after inflation to assess when the 
solution begins tracking. In the slow-roll approximation, the solution for the 
tracker is
\begin{equation}
\tilde{Q}(a)\simeq\(\frac{\beta(\beta+2)V_0}
{24\pi\rho_{\mathrm{f}0}(1+w_\mathrm{f})}\)^{1/(\beta+2)}
\(\frac{a}{a_0}\)^{3(1+w_\mathrm{f})/(\beta+2)}\comma
\end{equation}
where $\rho_{\mathrm{f}0}$ is the present value of the energy density and
$w_\mathrm{f}$ the pressure--density ratio of the dominant fluid (either
radiation or non-relativistic matter). Some examples of trackers are plotted in 
\reffig{tracker}. After inflation the quintessence field remains constant as 
long as its energy density is lower than the tracker's. We can therefore 
find the probability distribution $f_z$ for $\log_{10}(z_\mathrm{tr}+1)$, where 
$z_\mathrm{tr}$ is the redshift at which the quintessence reaches the tracking 
behaviour, by comparing the value of $Q$ at the end of inflation with the 
tracking solution. Using \reffig{peak} and \reffig{tracker} we can estimate at 
which redshift the tracker is reached for some values of $\alpha$ and $\beta$. 
We notice that the smaller $\beta$ is, the later the tracker is reached for a 
given initial value of $Q$. In \reffig{redshift} we show some examples of 
distributions for the redshift at which tracking begins. The discontinuity in 
the distribution comes from the fact that at about radiation--matter equality 
there is a transition between two trackers.

The striking feature of this figure is how low the expected redshift of tracking 
is. While some previous papers have advocated equipartition of the quintessence 
energy density as an initial condition \cite{equip}, leading to prompt tracking, 
we find that tracking is postponed until the late stages, and in particular well 
after nucleosynthesis. Indeed, the bulk of the probability is not only after 
nucleosynthesis but after decoupling too; however there is only a small 
probability that tracking has not begun by the present, which would not lead to 
acceptable quintessence.

Finally we must say that our results are valid for small initial values of the 
quintessence field before inflation (in fact formally zero). If this were not 
the case, the probability distribution for the mean value of the quintessence 
field  at the end of inflation will broaden to larger values and the tracker 
will be reached even later or not at all.

\section{Conclusions}

We have analyzed the quantum dynamics of the quintessence field during a period 
of early Universe inflation. Due to quantum fluctuations, even if the initial 
value of $Q$ in a certain region of the Universe is small, it is rapidly 
diffused to large field values and hence low energy density. We have found that 
typically tracking behaviour begins only at quite a late stage of evolution, 
well after nucleosynthesis and quite likely after decoupling too.

Although we have discussed specific models of both inflation and quintessence, 
we expect our results to be quite general; as far as inflation is concerned we 
need only the assumption that there are significantly more than fifty 
$e$-foldings 
in total and a standard value of $H$ during inflation, while for other models of 
quintessence where the potential diverges at the origin the result should also 
be the same as the precise form of the potential is dynamically irrelevant. In 
particular, while we have only considered models where the tracking density 
during nucleosynthesis is negligible, these considerations may reinstate models 
whose tracking density would be unacceptable during nucleosynthesis.


\begin{acknowledgments}
M.M.~was supported by the Fondation Barbour, the Fondation Wilsdorf and the 
Janggen-P\"{o}hn-Stiftung, and A.R.L.~in part by the Leverhulme Trust. We thank 
Pier-Stefano Corasaniti, Ruth Durrer, Anne Green, Lev Kofman, Dmitri Pogosyan 
and Emmanuel Zabey for useful discussions.
\end{acknowledgments}



\begin{thebibliography}{}
\bibitem{track} C. Wetterich, Nucl. Phys. {\bf B302}, 668 (1988).   
\bibitem{RP} B. Ratra and P. J. E. Peebles, Phys. Rev. D{\bf 37}, 3406
	(1988).
\bibitem{nucl2} E. J. Copeland, A. R. Liddle, and D. Wands, Ann. N. Y. Acad.
	Sci. {\bf 688}, 647 (1993); P. G. Ferreira and M. Joyce, Phys. Rev. 
	Lett. {\bf 79}, 4740 (1997), {\tt astro-ph/9707286}, Phys. Rev D 
	{\bf 58}, 023503 (1998), {\tt astro-ph/9711102}; M. Yahiro, G. J.
	Mathews, K. Ichiki, T. Kajino, and M. Orito, Phys. Rev. D{\bf 65},
	063502 (2002), {\tt astro-ph/0106349}.
\bibitem{scal}  E. J. Copeland, A. R. Liddle, and D. Wands, Phys. Rev. 
	D{\bf 57}, 4686 (1998), {\tt gr-qc/9711068}; A. R. Liddle and 
	R. J. Scherrer, Phys. Rev. D{\bf 59}, 023509 (1999), {\tt
	astro-ph/9809272}. 
\bibitem{equip} I. Zlatev, L. Wang, and P. J. Steinhardt, Phys. Rev. Lett. 
	{\bf 82}, 896 (1999), {\tt astro-ph/9807002}; P. J. Steinhardt, 
	L. Wang, and I. Zlatev, Phys. Rev. D{\bf 59}, 123504 (1999), 
	{\tt astro-ph/9812313}.
\bibitem{nucleo} R. Bean, S. H. Hansen and A. Melchiorri, Phys. Rev. D{\bf 64},
	103508 (2001), {\tt astro-ph/0104162}.
\bibitem{quintinf} P. J. E. Peebles and A. Vilenkin, Phys. Rev. D{\bf 59},
	063505 (1999), {\tt astro-ph/9810509}; E. J. Copeland, A. R. Liddle,
	and J. E. Lidsey, Phys. Rev. D{\bf 64}, 023509 (2001), {\tt
	astro-ph/0006421}; G. Huey and J. E. Lidsey, Phys. Lett. B{\bf 514},
	217 (2001), {\tt astro-ph/0104006}.
\bibitem{obscon} P. Brax, J. Martin, and A. Riazuelo, Phys. Rev. D{\bf 62},
	103505 (2000), {\tt astro-ph/0005428}; M. Doran, M. Lilley, and C.
	Wetterich, Phys. Lett. B{\bf 528},
	175 (2002), {\tt astro-ph/0105457}; P. S. Corasaniti and E. J. 
	Copeland, Phys. Rev. {\bf D65}, 043004 (2002), {\tt astro-ph/0107378};
	C. Baccigalupi, A. Balbi, S. Matarrese, F. Perrotta, and N. Vittorio,
	Phys. Rev. D{\bf 65}, 063520 (2002), {\tt astro-ph/0109097}
\bibitem{BHM02} R. Bean, S. H. Hansen, and A. Melchiorri, {\tt
	astro-ph/0201127}.
\bibitem{HKP} W. L. Freedman et al., Astrophys. J. {\bf 553}, 47 (2001),
	{\tt astro-ph/0012376}.
\bibitem{LL} A. R. Liddle and D. H. Lyth, {\em Cosmological Inflation
	and Large-Scale Structure}, Cambridge University Press,
	Cambridge, 2000.
\bibitem{SI} A. Vilenkin, Phys. Rev. D{\bf 27}, 2848 (1983); A. D. Linde, Phys.
	Lett. B{\bf 175}, 395 (1986).
\bibitem{SS} V. Sahni and A. Starobinsky, Int. J. Mod. Phys. {\bf D9}, 373
	(2000), {\tt astro-ph/9904398}.
\bibitem{SI2} A. D. Linde and A. Mezhlumian, Phys. Lett. B{\bf 307}, 25 (1993),
	{\tt gr-qc/9304015}; A. D. Linde, D. Linde, and A. Mezhlumian, Phys. 
	Rev. D{\bf 49},	1783 (1994), {\tt gr-qc/9306035}.
\end{thebibliography}
\end{document}